\documentclass[review,authoryear,12pt]{elsarticle}

\usepackage{graphicx}
\usepackage[tight,footnotesize, nooneline]{subfigure}

\usepackage{adjustbox} 
\usepackage[hidelinks]{hyperref} 

\usepackage{amssymb}
\usepackage{amsthm}
\usepackage{amsmath}

\usepackage{lineno}
 
\usepackage{multirow}

\usepackage[flushleft]{threeparttable} 

\journal{arXiv}

\newcommand\copyrighttext{%
   \textcopyright\ 2019 Springer Nature Limited. This manuscript version is made available under the CC-BY 4.0 license. \url{https://creativecommons.org/licenses/by/4.0/}}

\begin{document}

\begin{frontmatter}



\title{Comprehensive feature selection for classifying the treatment outcome of high-intensity ultrasound therapy in uterine fibroids}


\author[Affil1]{Visa Suomi \corref{cor1}}
\author[Affil1]{Gaber Komar}
\author[Affil2]{Teija Sainio}
\author[Affil1]{Kirsi Joronen}
\author[Affil1]{Antti Perheentupa}
\author[Affil1]{Roberto Blanco Sequeiros}

\address[Affil1]{Department of Radiology, Turku University Hospital, Kiinamyllynkatu 4-8, 20521 Turku, Finland}
\address[Affil2]{Department of Medical Physics, Turku University Hospital, Kiinamyllynkatu 4-8, 20521 Turku, Finland}
\cortext[cor1]{Corresponding Author: Visa Suomi, Department of Radiology, Turku University Hospital, Kiinamyllynkatu 4-8, 20521 Turku, Finland; Email, visa.suomi@tyks.fi}

\begin{abstract}
The study aim was to utilise multiple feature selection methods in order to select the most important parameters from clinical patient data for high-intensity focused ultrasound (HIFU) treatment outcome classification in uterine fibroids. The study was retrospective using patient data from 66 HIFU treatments with 89 uterine fibroids. A total of 39 features were extracted from the patient data and 14 different filter-based feature selection methods were used to select the most informative features. The selected features were then used in a support vector classification (SVC) model to evaluate the performance of these parameters in predicting HIFU therapy outcome. The therapy outcome was defined as non-perfused volume (NPV) ratio in three classes: $<$30\%, 30-80\% or $>$80\%. The ten most highly ranked features in order were: fibroid diameter, subcutaneous fat thickness, fibroid volume, fibroid distance, Funaki type I, fundus location, gravidity, Funaki type III, submucosal fibroid type and urinary symptoms. The maximum F1-micro classification score was 0.63 using the top ten features from Mutual Information Maximisation (MIM) and Joint Mutual Information (JMI) feature selection methods. Classification performance of HIFU therapy outcome prediction in uterine fibroids is highly dependent on the chosen feature set which should be determined prior using different classifiers.
\end{abstract}


\end{frontmatter}

\copyrighttext

\pagebreak



\section*{Introduction}

Uterine fibroids (aka myomas) are benign tumours of the uterus which are formed by the excessive growth of smooth-muscle cells in the wall of the uterus~\citep{stewart2001uterine}. They tend to be round-shaped with well-defined boundaries and their diameter can range from 1~cm to more than 10~cm \citep{walker2002uterine}. These features make their diagnosis relatively easy using ultrasound (US) or magnetic resonance imaging (MRI) \citep{dueholm2002accuracy}. They are commonly diagnosed in all women, but certain factors such as age, ethnicity and heredity increase the risk~\citep{sparic2016epidemiology, stewart2017epidemiology}. Uterine fibroids occur in up to 70\% of women, which makes them the most common neoplasm affecting women~\citep{stewart2017epidemiology}.

Uterine fibroids are asymptomatic in the majority of cases, but they may occasionally cause symptoms such as abnormal uterine bleeding, pelvic pressure and pain, and reproductive dysfunction~\citep{stewart2001uterine, zimmermann2012prevalence}. Uterine fibroids can be treated using surgical methods such as hysterectomy or open/laparoscopic myomectomy, and in some cases hormonal therapy or embolisation might also be suitable. Moreover, in recent years alternative therapy methods such as high-intensity focused ultrasound (HIFU) have gained popularity. The benefits of HIFU therapy are its complete non-invasiveness, lower costs related to patient recovery time and low complication rate~\citep{kim2011mr}. However, not all patients diagnosed with uterine fibroids are suitable candidates for HIFU treatment, which currently limits the success rate of the therapy outcome.

In clinical practice suitable patients for the HIFU therapy are typically screened by a radiologist together with a gynaecologist~\citep{kim2014techniques}. For this purpose, evaluation criteria have been formed for patient selection~\citep{funaki2009clinical, wei2018predictive, keserci2018magnetic}, but these are usually only based on MR images, and hence, lack more comprehensive inclusion of clinical aspects such as the treatment history, symptoms and patient-specific physiology. Therefore, more refined models are needed to evaluate the suitability of patients for HIFU treatments and to predict the treatment outcome. In order to build these prediction models, all the clinical and patient parameters affecting the treatment outcome need to be first comprehensively and rigorously determined.

The aim of this study was to utilise different filter-based feature selection methods to identify the most important features from clinical patient data for HIFU therapy outcome classification in uterine fibroids. More specifically, to answer the questions: 
\begin{enumerate}
\item Which features are the most informative in classifying the treatment outcome?
\item What is the optimal number of features to achieve the best classification performance?
\item Are there any specific feature selection methods that perform better in choosing these features?
\end{enumerate}
In order to answer these questions, the features chosen by the feature selection methods were used in a support vector classification (SVC) model to evaluate their performance in treatment outcome prediction. The results of this study identify the most prominent clinical parameters for HIFU therapy outcome prediction in uterine fibroids, which will advance the development and adaptation of these models into clinical practice.

\section*{Methods}

\subsection*{Data collection}

The study was retrospective using the patient data from the uterine fibroid HIFU treatments conducted at the Turku University Hospital, Finland, between May 2016 and January 2019. To be eligible for inclusion, the patients had to be treated for at least one uterine fibroid with Sonalleve V2 (Profound Medical, Mississauga, Canada) therapy system. A total number of 66 patients with 89 uterine fibroids were included in the study using these criteria (see Table~\ref{tab:patient_data}). Ethical permission for the usage of the patient data (ETMK: 152/1801/2016) was obtained before the study from the Ethics Committee of Hospital District of Southwest Finland. All methods were carried out in accordance with relevant guidelines and regulations and all experimental protocols were approved by the committee. Informed consent was obtained from all subjects and all health records collected for the study were anonymised prior using the data. All subjects were over 18 years old.

\subsection*{Class labelling}

The objective of the SVC model was to classify uterine fibroids according to their predicted HIFU treatment outcome. For this purpose, the treated uterine fibroids were divided into three classes based on their immediate post-treatment non-perfused volume (NPV) ratios, i.e., the ratio of the ablated volume in MR perfusion image compared to the total size of the fibroid. The NPV ratio was selected as the class label because it can be objectively determined and has been shown to correlate with the clinical treatment outcome of patients with uterine fibroids~\citep{fennessy2007uterine, mindjuk2015mri}. NPV ratio of more than 80\% has been shown to result in clinical success in more than 80\% of patients~\citep{mindjuk2015mri}. A clinical success in this case was defined as at least 10-point reduction in the symptom severity score after the treatment, which is considered as a common criterion for significant symptom improvement~\citep{stewart2006clinical}. On the other hand, NPV ratio of less than 30\% has been shown to have odds for clinical success of only one third against those with NPV ratio of 30\% or more~\citep{fennessy2007uterine}. Based on these thresholds, the class labels were defined as: `class 0' for poor outcome (NPV $<$ 30\%), `class 1' for mediocre outcome (30\% $\leq$ NPV $\leq$ 80\%), and `class 2' for good outcome (80\% $<$ NPV). Each treated uterine fibroid was categorised into one of these three classes (see Table~\ref{tab:patient_data}).

\begin{table}[t!]
  \centering
  \caption{Summary of study patients and treatment outcome class labels.}
    \begin{tabular}{lllllll}
    \hline
    \textbf{Category}				&		&		&		&		&		& \textbf{Count}	\\
    \hline
    Treated patients				&		&		&		&		&		& 66				\\
    Uterine fibroids				&		&		&		&		&		& 89				\\
    Non-perfused volume (NPV) ratio	&  		&  		& NPV	& $<$	& 30\%	& 15				\\
    								& 30\% 	& $\leq$& NPV 	& $\leq$& 80\%	& 52				\\
   									& 80\%	& $<$ 	& NPV	&		&		& 22				\\
	\hline
    \end{tabular}
  \label{tab:patient_data}
\end{table}

\subsection*{Feature extraction}

The features for therapy outcome classification were extracted manually from pre-treatment MRI scans as well as electronic medical records stored at the hospital. A total of 39 parameters covering different clinical aspects of the treatments were selected as features (see Tables~\ref{tab:numerical_features} and~\ref{tab:categorical_features} for numerical and categorical features, respectively). These features were selected based on their relevance to clinical HIFU treatments, the availability of the data in medical records and empirically observed correlations with the treatment outcome.

\begin{table}[t!]
  \centering
  \caption{Numerical features with their mean, standard deviation (SD), minimum and maximum values per fibroid.}
    \begin{tabular}{lllll}
    \hline
    \textbf{Feature} & \textbf{Mean} & \textbf{SD} & \textbf{Min} & \textbf{Max} \\
    \hline
    Age (years) 					& 41.9  & 6.0   & 26    & 51 \\
    Weight (kg) 					& 68.0  & 12.2  & 44    & 108 \\
    Height (cm) 					& 165.4 & 6.2   & 152   & 178 \\
    Gravidity 						& 1.2   & 1.5   & 0     & 7 \\
    Parity 							& 0.8   & 1.0   & 0     & 4 \\
    Subcutaneous fat thickness (mm) & 16.2  & 7.9   & 3.3   & 41.2 \\
    Front-back distance (mm) 		& 141.0 & 17.9  & 89.2  & 173.3 \\
    Fibroid diameter (mm) 			& 42.0  & 20.5  & 8.6   & 89.8 \\
    Fibroid distance (mm) 			& 47.5  & 18.7  & 15.6  & 91.8 \\
    Fibroid volume (ml) 			& 84.4  & 126.0 & 0.4   & 898 \\
    \hline
    \end{tabular}
  \label{tab:numerical_features}
\end{table}

\begin{table}[t!]
  \centering
  \caption{Categorical features with counts per fibroid.}
  	\begin{adjustbox}{totalheight=\textheight-2\baselineskip}
    \begin{tabular}{lll}
    \hline
   	\textbf{Feature} & \textbf{Categories} & \textbf{Count} \\
   	\hline
    \multicolumn{1}{l}{Ethnicity} & White & 80 \\
          & Black & 6 \\
          & Asian & 3 \\
    \multicolumn{1}{l}{Previous pregnancies} & Yes   & 50 \\
          & No    & 39 \\
    \multicolumn{1}{l}{Live births} & Yes   & 42 \\
          & No    & 47 \\
    \multicolumn{1}{l}{C-section} & Yes   & 4 \\
          & No    & 85 \\
    \multicolumn{1}{l}{Treatment history} & Esmya & 30 \\
          & Open myomectomy & 4 \\
          & Laparoscopic myomectomy & 3 \\
          & Hysteroscopic myomectomy & 14 \\
          & Embolisation & 1 \\
    \multicolumn{1}{l}{Abdominal scars} & Yes   & 24 \\
          & No    & 65 \\
    \multicolumn{1}{l}{Symptoms} & Bleeding & 69 \\
          & Pain  & 8 \\
          & Mass  & 17 \\
          & Urinary & 11 \\
          & Infertility & 15 \\
    \multicolumn{1}{l}{Fibroid type} & Intramural & 54 \\
          & Subserosal & 8 \\
          & Submucosal & 31 \\
    \multicolumn{1}{l}{Fibroid location} & Anterior & 44 \\
          & Posterior & 38 \\
          & Lateral & 31 \\
          & Fundus & 7 \\
    \multicolumn{1}{l}{Uterus position} & Anteverted & 80 \\
          & Retroverted & 9 \\
    \multicolumn{1}{l}{Funaki type} & Type I & 18 \\
          & Type II & 66 \\
          & Type III & 5 \\
    \hline
    \end{tabular}
    \end{adjustbox}
  \label{tab:categorical_features}
\end{table}

\subsection*{Feature selection methods}

A total of 14 different filter-based (i.e., independent of the classification model) feature selection methods were used in the analysis: Chi-square (CHI2), conditional infomax feature extraction (CIFE), conditional mutual information maximisation (CMIM), double input symmetric relevance (DISR), Fisher score (FISH), F-score (FSCR), Gini index (GINI), interaction capping (ICAP), joint mutual information (JMI), mutual information feature selection (MIFS), mutual information maximisation (MIM), minimum redundancy maximum relevance (MRMR), relief (RELF) and trace ratio (TRAC). In addition, the aggregate rankings from all feature selection methods will be referred as TOPN. All feature selection methods are publicly available from the scikit-feature open access repository~\citep{li2018feature}.

The advantage of using filter-based methods compared to wrapper and embedded methods are that they are computationally more efficient and less prone to overfitting~\citep{guyon2003introduction}. Filter-based methods rank the features using a scoring criterion and they can be divided into univariate or multivariate methods depending on whether their consider variables one-by-one or together in groups, respectively. In univariate methods, the scoring criterion only depends on the feature relevancy while the feature redundancy is ignored. In multivariate methods, the multivariate interaction within the features and the scoring criterion is a weighted sum of feature relevancy and redundancy. More detailed descriptions about these methods can be found from the reference~\citep{li2018feature}.

\subsection*{Treatment outcome classification and hyperparameter search}

\begin{figure}[b!]
\centering
\includegraphics[width=1\textwidth]{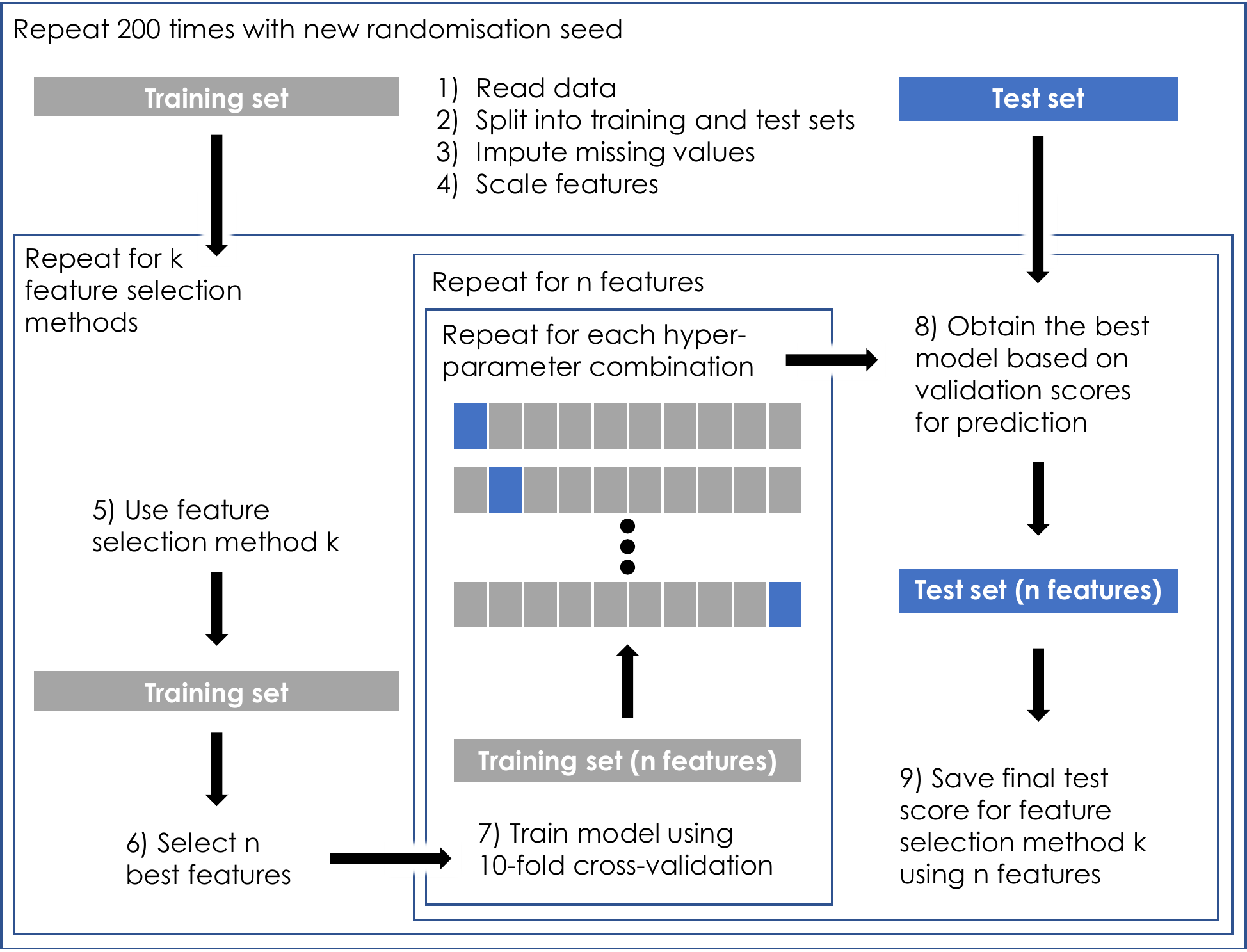}
\caption{Overview of the data processing pipeline: 1) The data were read to a dataframe; 2) split into training and test sets; 3) imputed with mean or mode values based on the training set; 4) the features were log-scaled; 5) feature selection method ($k$ = 1-14) was used on the training set; 6) the highest-ranking features ($n$ = 2-20) were obtained; 7) the $n$ features from method $k$ were used to train a support vector classification (SVC) model using hyperparameter grid search with inner 10-fold cross-validation; 8) the SVC model was refit on the whole training set using the combination of hyperparameters based on the highest cross-validation score (F1-micro); 9) the fitted SVC model was used to classify uterine fibroids in the test set with the same $n$ features and the test score (F1-micro) was saved. Steps 1-9) were repeated 200 times with a new randomisation seed.}
\label{fig:algorithm_pipeline}
\end{figure}

An overview of the data processing pipeline is presented in Figure~\ref{fig:algorithm_pipeline}. The dataset of 89 fibroids was first read to a dataframe and was randomly divided into training and test sets with 71:18 ratio, respectively. The splitting was stratified which ensured that the ratio of different classes was the same in both training and test data. Height and gravidity were the only variables with missing values (24.7\% and 7.9\% of values missing, respectively) and thus they were imputed using their mean and mode, respectively. The imputation values were based only on the training data in order to avoid any information leakage from the test data. 

The categorical features were one-hot encoded and all features were logarithmically scaled in order to optimise the classification model performance. Since the treatment outcome class distribution was unbalanced, class weights were also calculated for each class in the training set in order to balance the classifier fitting process. Once the data were pre-processed, the feature selection methods ($k$ = 1-14) were used on the training set to select the best features ($n$ = 2-20). These $n$ features from the training set were then used as inputs for the classification model.

SVC model was selected as the classification method for the problem since it does not incorporate any built-in feature selection methods (i.e., embedded or wrapper) and because of its ability to perform well with small but clean datasets. SVC is a supervised machine learning model whose basic idea is to classify data into correct classes by using a set of hyperplanes. These hyperplanes can either be linear or nonlinear depending on the kernel of the model. In this study, nonlinear radial basis function (RBF) kernel was used. The SVC model was implemented using scikit-learn (v0.20.2) in Python (3.7.2)~\citep{pedregosa2011scikit}.

Before starting the SVC model fitting process with the selected $n$ features from the training data, a set of hyperparameters were determined for grid search (see Table~\ref{tab:hyperparameters}). The aim of the grid search was to find the optimal combination of hyperparameters for each set of features by iterating over all the possible combinations of the given grid. The performance of each hyperparameter combination was evaluated using stratified 10-fold cross-validation on the training data and calculating a validation score (F1-micro) for each fold. The mean score over all 10 folds was then selected as the performance metric of the given hyperparameter combination and the process was repeated for the next set of hyperparameters. Once all hyperparameter combinations had been evaluated, the best mean score was selected as the optimal set of hyperparameters for the given $n$ features. The SVC model was then refitted on the whole training set using these hyperparameters. Finally, the fitted model was used to make predictions on the test set with the same $n$ features and the test score (F1-micro) was calculated based on the test predictions as the final performance metric.

The whole process above was repeated 200 times using a new randomisation seed in every iteration, which was found to ensure the stability and repeatability of the results. All the results are thus average values over 200 iterations. The total computation time was about 10 hours using a desktop computer with CPU processing (Intel Xeon E5-2643 v3 @ 3.40~GHz).

\begin{table}[t!]
  \centering
  \caption{Hyperparameters for grid search in support vector classification (SVC) model fitting.}
    \begin{tabular}{ll}
    \hline
    Parameter & Values  \\
    \hline
    Kernel 			& RBF \\
    C     			& 1e-1, 1, 1e1, 1e2, 1e3, 1e4 \\
    Gamma 			& 1e-2, 1e-1, 1, 1e1, 1e2, 1e3, 1e4 \\
    \hline
    \end{tabular}
  \label{tab:hyperparameters}
\end{table}

\section*{Results}

\subsection*{Feature importance and correlation}

\begin{figure}[b!]
\centering
\includegraphics[width=1\columnwidth]{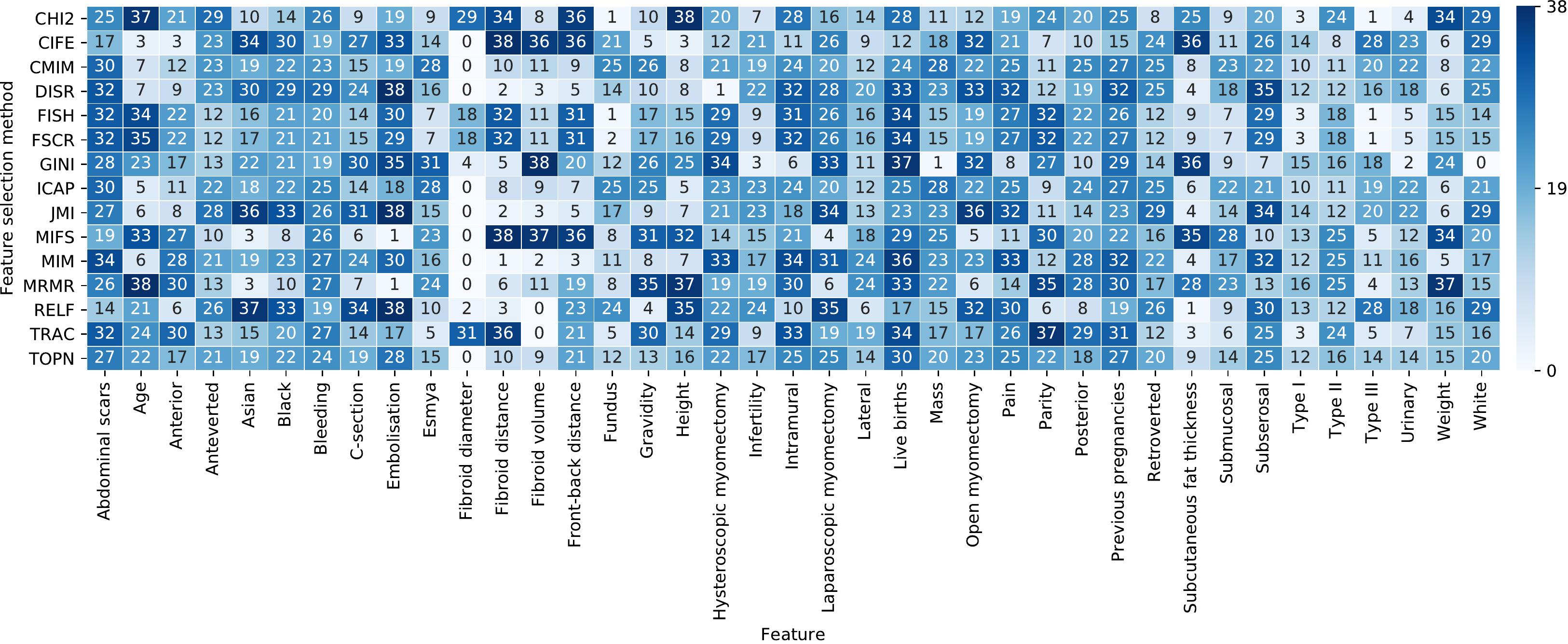}
\caption{Heatmap showing the median feature rankings (range 0-38 from the best to worst) from 14 different filter-based feature selection methods (200 rankings per feature) in classifying HIFU treatment outcome in uterine fibroids. In addition, the median ranking from all methods (14 methods $\times$ 200 repetitions = 2800 rankings per feature) for each feature is shown at the bottom (TOPN).}
\label{fig:heatmap_rankings_median}
\end{figure}

A heatmap of the median feature rankings (range 0-38 from the best to worst) in classifying HIFU treatment outcome in uterine fibroids is shown in Figure~\ref{fig:heatmap_rankings_median}. The rankings are shown for all 39 features from the 14 different feature selection methods. Each feature had 200 rankings per method and their corresponding median value is presented in the map. In addition, the aggregate median rankings from all methods (14 methods $\times$ 200 repetitions = 2800 rankings per feature) are shown at the bottom (TOPN).

It is evident from Figure~\ref{fig:heatmap_rankings_median} that most methods (9 out of 14) are rather unanimous about fibroid diameter being the most informative feature in classifying the HIFU treatment outcome in terms of the immediate post-treatment NPV ratio. This is somewhat expected since large fibroids take longer time to treat, and with limited treatment time windows, they cannot always be fully ablated. Furthermore, due to their large size some parts of the fibroid might be out of the reach of the ultrasound focal point or they might be located behind adjacent tissue structures, such as the bowel, preventing treatment delivery to these locations. It is also noticeable that some methods have strong disagreement about the rankings. FISH and FSCR rank fibroid diameter approximately in the middle (18th) while CHI2 and TRAC place it near the end (29th and 31st, respectively).

\begin{figure}[b!]
\centering
\includegraphics[width=1\columnwidth]{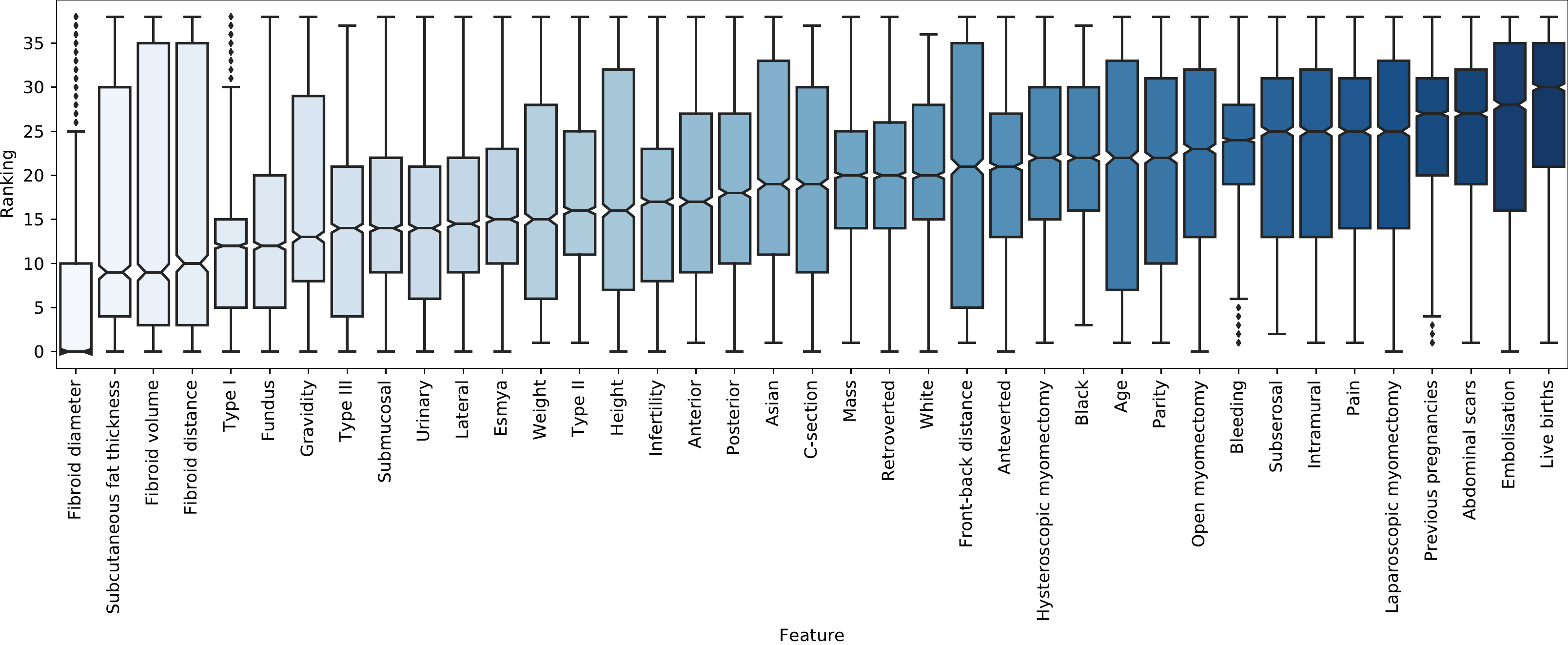}
\caption{Boxplot showing the feature rankings (range 0-38 from best to worst) from 14 different filter-based feature selection methods (14 methods $\times$ 200 repetitions = 2800 rankings per feature) in classifying HIFU treatment outcome in uterine fibroids. The features are ordered by their median value based on the rankings. Boxes show the interquartile ranges (IQR) with median values (notch) and whiskers show 1.5 IQR from the lower and upper quartiles. Outliers are plotted as individual points beyond the ends of the whiskers.}
\label{fig:boxplot_feature_rankings}
\end{figure}

\begin{figure}[b!]
\centering
\includegraphics[height=6cm]{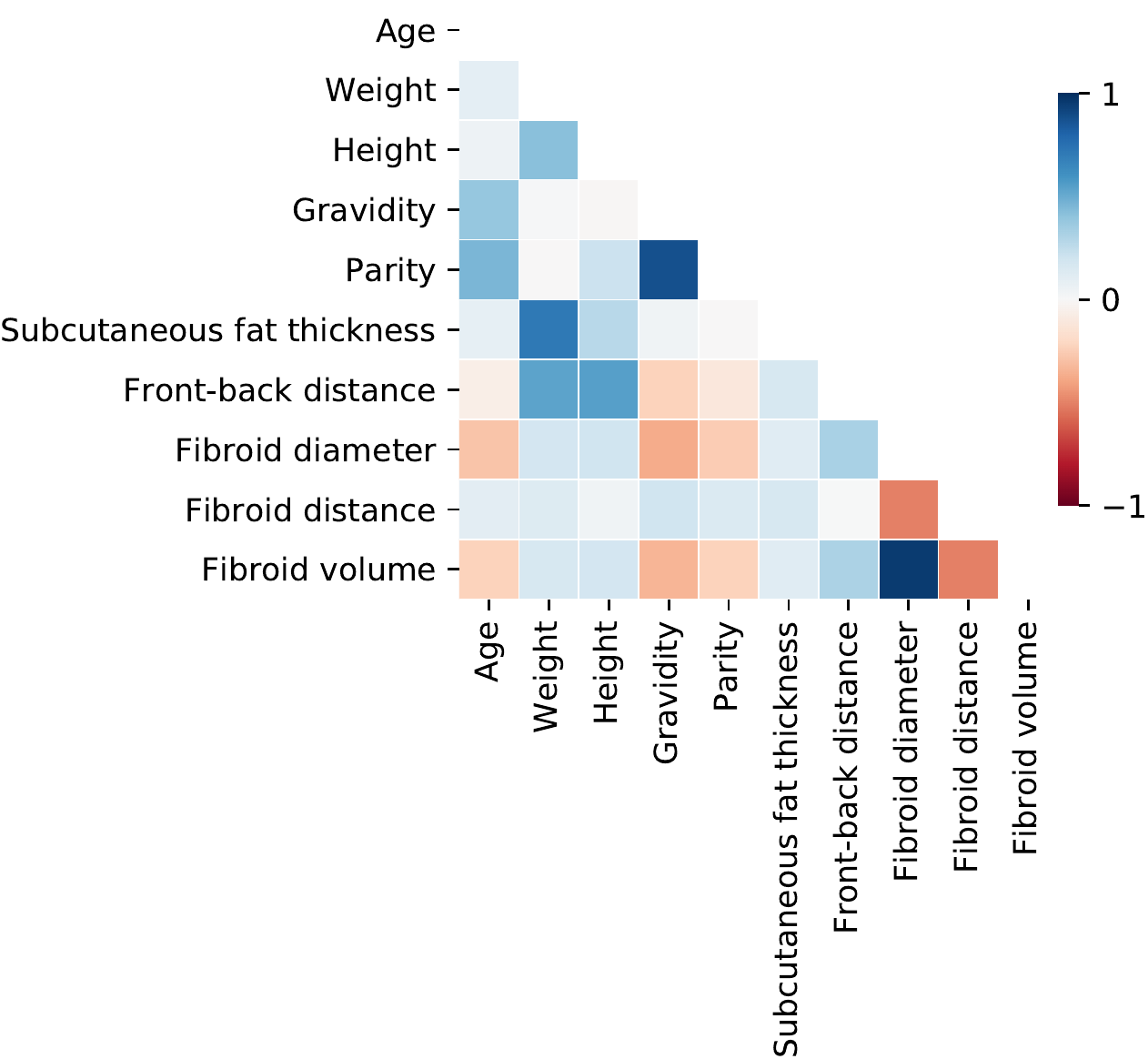}
\caption{Diagonal correlation matrix showing the pairwise Spearman's correlation coefficients between numerical features.}
\label{fig:feature_corr}
\end{figure}

The statistical distributions of the aggregate rankings are visualised in the boxplot in Figure~\ref{fig:boxplot_feature_rankings} which shows the features ordered by their median rankings (i.e., TOPN in Figure~\ref{fig:heatmap_rankings_median}). The boxes display the interquartile ranges (IQR) and the median values are marked with a notch inside each box. The whiskers show 1.5 IQR from the lower and upper quartiles and outliers are plotted as individual points beyond the ends of the whiskers. In addition, the pairwise Spearman's correlation coefficients between numerical features are shown in Figure~\ref{fig:feature_corr}.

The ten highest-ranking features organised by their aggregate median value were: 1) fibroid diameter, 2) subcutaneous fat layer thickness, 3) fibroid volume, 4) fibroid distance from the skin surface, 5) Funaki type I, 6) fundus location, 7) gravidity, 8) Funaki type III, 9) submucosal fibroid type and 10) urinary symptoms. Some of the features have relatively large IQRs and whiskers, which indicates a certain degree of disagreement between the methods. There are also visible outliers outside the whiskers in some of the features. Nevertheless, it is evident that features related to the physical measures of the fibroid and patient geometry are ranked highest. It should also be noted that Figure~\ref{fig:boxplot_feature_rankings} shows the highest-ranking features from the aggregate votes, and thus, they may not represent the optimal feature set for the best treatment outcome classification performance as will be seen in the next section.

Fibroid diameter and fibroid volume ranked first and third, respectively, both of which indicate the size of the fibroid. It is obvious that both of these features are also strongly correlated (see Figure~\ref{fig:feature_corr}), but diameter seems to contain more information in terms of outcome classification when compared to fibroid volume. As mentioned earlier, larger fibroids take longer to treat, and hence, the treatment time constraint limits their final NPV ratio after the treatment. Furthermore, parts of a large fibroid might also be located outside the allowed sonication region by the system or they might be located behind adjacent tissue structures preventing the treatment delivery.

Subcutaneous fat layer thickness ranked second overall despite of its relatively large statistical dispersion. Ultrasound energy is strongly attenuated by the subcutaneous fat layer, and thus, a thick fat layer efficiently reduces heating efficacy of the ultrasound field. Poor heating efficacy consequently translates to lower expected NPV ratio. Furthermore, the thickness of fat layer affects the cooling times between sonications, which causes the total treatment time become longer. Subcutaneous fat layer thickness is also strongly correlated with patient weight (ranked 13th) as seen in Figure~\ref{fig:feature_corr}, but since weight is also affected by height (ranked 15th) and muscle mass, it is not a direct indication of the fat content \textit{per se}. For this reason, subcutaneous fat layer thickness contains more predictive information in this case.

Fibroid distance from the skin surface was ranked fourth overall. Longer propagation distance results in higher attenuation of the ultrasound energy, which has fundamentally the same effect as the subcutaneous fat layer. Fibroid distance also adds another element to this measurement by providing information about the intra-abdominal fat which varies significantly and is not necessarily in correlation with the subcutaneous fat layer. Furthermore, longer distance affects the reachability of the fibroid since the therapeutic transducer has a fixed focal length (14~cm) from its surface. Fibroids that are located very far back have therefore worse prospective treatment outcome due to the combined effect of high ultrasound attenuation and limited reachability. It should be noted that the usage of different bowel manipulation techniques~\citep{jeong2017usefulness} might affect this distance. However, the distances in this study were measured from treatment planning images, which should correspond to the actual distances during the treatments.

The feature quantifying the distance between the skeletal muscle and the back-bone (i.e., front-back distance) ranked 24th. In theory, the smaller the front-back distance the easier it is to manipulate the location of the myoma using different gel-filling techniques. However, this parameter did not seem contain much predictive information about the treatment outcome.

Funaki types are clinically used categories to classify the intensity of uterine fibroid based on T2-weighted MR images. The fibroids are divided into three Funaki types according to their average T2-signal intensity compared to those of the myometrium and skeletal muscle. The high-intensity fibroids on the T2-weighted images have been shown to represent high vascularisation, fluid-rich tissues or degeneration~\citep{funaki2007magnetic}, which generally results in diminished heating efficacy. By their definition, Funaki types I (low T2 intensity) and III (high T2 intensity) predict good and poor therapy outcome, respectively. They were ranked fifth and eighth overall, which further proposes their ability to contain information about the treatment outcome in terms of the NPV ratio. Funaki type II ranked 14th, and thus, it is not as informative predictor for treatment outcome as the other two.

Fundus location (i.e., fibroids located at the top of the uterus) was ranked sixth which most likely relates to its good reachability during HIFU treatments. This is especially true when the uterus is in anteverted position, which is the most common case (see Table~\ref{tab:categorical_features}). Fibroids located at the top of the uterus are typically easier to reach with the limited coverage of the ultrasound field, and therefore, their prospective treatment outcomes are usually good. The other locations (lateral, anterior and posterior) do not seem to be quite as informative with the respective rankings of 11th, 17th and 18th. Retroverted and anteverted uterus positions ranked 22nd and 25th, respectively. Anteverted position is typically more desirable for the reachability of fibroids, but its predictive information content is slightly lower compared to retroverted position.

Gravidity, specifying the number of pregnancies, ranked seventh overall. It is not completely clear why this number would have predictive information about the therapy outcome since previous/undergone pregnancies do not affect the hormonal levels in women. However, it might be related to the size of the uterus after multiple pregnancies. Another possible explanation is that multiple pregnancies could affect the type or location of the fibroids. There was also some disagreement about its ranking between different feature selection methods (i.e., relatively large IQR). In Figure~\ref{fig:feature_corr} it is shown that gravidity is strongly correlated with parity (i.e., the number of live births), which was ranked only 29th overall. The same features as Boolean variables (i.e., previous pregnancies and live births) ranked 36th and 39th overall. Therefore, the Boolean parameters are not as informative as their corresponding numerical values.

Fibroid types are determined by their growth direction with submucosal fibroids growing towards the uterine cavity while subserosal fibroids project to the outside of the uterus and intramural fibroids grow within the muscular uterine wall. It should be noted that these types are not mutually exclusive and hence some fibroids might be of more than one type. Submucosal fibroids were ranked ninth overall with subserosal and intramural fibroids ranking 32th and 33th, respectively. Submucosal fibroids are usually also the most problematic in terms of their symptoms with heavy bleeding and they have often been previously treated with hysteroscopic myomectomy due to the same reason. One hypothesis might be that these previous interventions have changed the internal tissue structure of the fibroids (in terms of vascularisation and fluid content), which also affects the therapy outcome. In addition, previous treatments may have reduced their size considerably, which is likely to result in better HIFU treatment outcome.

Urinary symptoms (i.e., incontinence) was ranked tenth overall. Fibroids causing urinary symptoms are typically located adjacent to bladder, which is a favourable location in terms of therapy efficacy. Ultrasound effectively penetrates through bladder consisting mainly of water, and thus, the energy reduction due to attenuation is minimal. This allows high heating efficacy at the target location, which often times results in high NPV ratio. The other symptoms (infertility, mass effect, bleeding and pain) ranked 16th, 21st, 31st and 34th, respectively, and were not considered to be as informative.

Previous treatments for uterine fibroids did not contain enough information in estimating the treatment outcome to be ranked in top ten. Hormonal therapy (i.e., esmya), hysteroscopic myomectomy, open myomectomy, laparoscopic myomectomy and embolisation ranked 12th, 26th, 30th, 35th and 38th, respectively. It has been shown that some hormonal therapies for uterine fibroids change the morphological tissue properties and vascular structure of the myoma~\citep{crow1995morphological, smart2006magnetic}, which would also affect its acoustic and thermal properties.

Likewise, ethnicity did not have a lot of predictive information related to the treatment outcome in the study population. The patient group consisted of Asian, white and black populations which ranked 19th, 23rd and 27th, respectively. It is known that black people are more likely to develop myomas~\citep{stewart2017epidemiology}, which might affect the prospective treatment outcome in terms of the size and number of fibroids. However, the large majority of the patient group were white (see Table~\ref{tab:categorical_features}) so this result might be somewhat biased, and thus, the importance of ethnicity was not reflected in the results.

Previous surgeries in the abdominal area also did not have considerable effect on the treatment outcome. The respective rankings C-section and the presence of abdominal scars were 20th and 37th. Other abdominal interventions such as laparoscopic or open myomectomy also cause scarring. Scars tend to heat considerably during therapeutic ultrasound treatments which is why they are usually covered using scar tape during the therapy. The scar tape causes part of the ultrasound energy be reflected which reduces the heating efficacy at the target location. However, this effect did not seem to be remarkable in terms of the treatment outcome.

Age ranked 28th overall and is known to affect the hormonal levels in women. All women in the study were premenopausal with regular menstrual cycle. The oestrogen and progesterone levels in blood have some cyclic variations but they do not usually differ markedly between women or change within the fertile age range~\citep{melmed2015williams}. The hormonal milieu within the fibroids may differ from that in blood circulation and be of more importance regarding the response of the fibroid to different medical therapies. The response of fibroids to hormonal stimuli may also vary depending on the structure of the fibroid, which is something that can be assessed with MRI. However, it should be noted that the patient population in the study was pre-screened using MRI with fibroids initially found suitable for HIFU treatment, thereby resulting in a more homogeneous fibroid group. Currently there are no data on association of hormonal levels and fibroid MRI appearance. Moreover, there are no data suggesting that levels of sex hormones have an effect on the response of the fibroid to HIFU therapy, which is also reflected in the results. Obtaining hormonal levels in fibroids would require an invasive biopsy which is not suitable in standard clinical practice.

\subsection*{Classification performance}

\begin{figure}[b!]
\centering
\includegraphics[width=1\columnwidth]{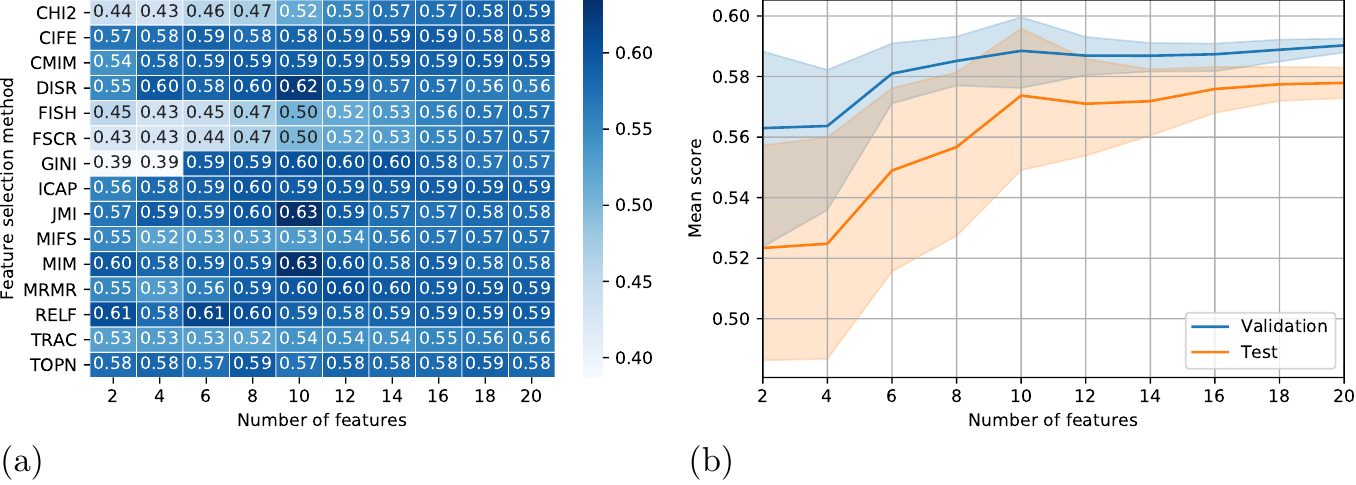}
\caption{(a) Heatmap showing the mean test scores (F1-micro) from support vector classification (SVC) model in classifying HIFU treatment outcome in uterine fibroids. The values show the mean test score from 200 repetitions using the stated number of highest-ranking features from each feature selection method. In addition, the mean test scores using the highest-ranking features from aggregate votes are shown at the bottom (TOPN). (b) Lineplot showing the mean and dispersion of validation and test scores with the number of highest-ranking features from all feature selection methods. The faded areas show the 95\% confidence intervals.}
\label{fig:cross-validation_scores}
\end{figure}


Once the most important features were determined using all the feature selection methods, their performance in classifying the treatment outcome was evaluated with the SVC model. For this purpose, the number of input features for the model was varied between 2 and 20 using the highest-ranking features from each method. Figure~\ref{fig:cross-validation_scores}(a) shows a heatmap of the average F1-micro scores over 200 iterations using the specified number of highest-ranking features from each feature selection method. In addition, the mean test scores using the highest-ranking features from aggregate votes (see Figure~\ref{fig:boxplot_feature_rankings}) are shown at the bottom (TOPN). Figure~\ref{fig:cross-validation_scores}(b) shows the mean and dispersion of the validation and test scores as a function of highest-ranking features from all feature selection methods. The faded areas show the 95\% confidence intervals.

In Figure~\ref{fig:cross-validation_scores}(a), most feature selection methods achieved relatively high performance scores with only two of the highest-ranking features. The best classification performance with two features was achieved using RELF whose mean test score was 0.61. The two highest-ranking features by their median value from RELF were fibroid volume and subcutaneous fat layer thickness (see Figure~\ref{fig:heatmap_rankings_median}). The worst performance was achieved using GINI whose mean test score with two features was 0.39. The two highest-ranking features from GINI were white ethnicity and mass symptom which clearly did not contain as much predictive information. Relatively low mean scores were also displayed by FSCR, CHI2 and FISH whose respective test scores were 0.43, 0.44 and 0.45. All of them used Funaki type III and fundus location as the two most important features.

From Figure~\ref{fig:cross-validation_scores}(b) it is evident that increasing the number of features improves the performance of the classification model. The validation and test scores increased continuously and peaked with ten features after which the scores stayed almost flat. The dispersions of both validation and test scores also got smaller when the number of features increased, which is a result of more robust classification performance. Hence, using the ten highest-ranking features seems like the optimum amount for the best classification performance with the SVC model.

In Figure~\ref{fig:cross-validation_scores}(a) the best classification performance was achieved using the ten highest-ranking features from MIM and JMI both of which achieved the mean test score of 0.63. The ten highest-ranking features from MIM in order were (see Figure~\ref{fig:heatmap_rankings_median}): 1) fibroid diameter, 2) fibroid distance, 3) fibroid volume, 4) front-back distance, 5) subcutaneous fat layer thickness, 6) weight, 7) age, 8) height, 9) gravidity and 10-11) Funaki type III/fundus location. Likewise, for JMI these were: 1) fibroid diameter, 2) fibroid distance, 3) fibroid volume, 4) subcutaneous fat layer thickness, 5) front-back distance, 6-7) age/weight, 8) height, 9) anterior location and 10) gravidity. The top eight features from both methods were the same (although not in the same order) after which there was some discrepancy.

When using the highest-ranking features from the aggregate median rankings (TOPN) in Figure~\ref{fig:boxplot_feature_rankings}, the performance of the classification model was actually slightly lower compared to the features chosen by some of the feature selection methods. This is because the aggregate rankings represent the median vote from all the methods, and therefore, their performance should also be around average. Thus, in order to achieve the best classification performance, different feature selection methods should be evaluated individually in order to determine their suitability for the chosen classification model. In this study, RELF, JMI and MIM were found to be the most ideal for the SVC model.

It should also be considered that some of these features are strongly correlated with each other (see Figure~\ref{fig:feature_corr}), and thus, using them at the same time does not necessary bring any additional predictive information to the model. Furthermore, collinearity between features might affect the predictive performance of the chosen model when they are used at the same time. For instance, fibroid diameter and fibroid volume are both indicators of the size of the myoma and the latter is just a volumetric transformation of the former (i.e., assuming spherical fibroid shape). Therefore, only one of these features would be enough to bring the necessary information to the model. In clinical setting typically only the fibroid diameter is readily available and thus it is preferable to the fibroid volume.

\section*{Discussion}

Previous research has shown that certain physiological and MRI parameters have significant statistical correlation with the HIFU treatment outcome in uterine fibroids~\citep{funaki2007magnetic, mindjuk2015mri, keserci2018magnetic}. \citet{funaki2007magnetic} divided the intensities in T2-weighted MR images of uterine fibroids into three separate classes (i.e., type I, II and III) using threshold values based on the skeletal muscle and myometrium. It was shown that the post-treatment NPV ratio correlated with these three classes and that type III fibroids had significantly lower NPV ratio compared to type I and II fibroids. Therefore, it was suggested that type I and II fibroids were suitable candidates for HIFU treatment whereas type III fibroids were not. Since then, this relatively simple classification method has been widely adopted in clinical practice to evaluate the treatability of uterine fibroids. The results in this study further suggest that at least Funaki types I and III have predictive information related to treatment outcome. However, the result might be slightly biased since type III fibroids were usually excluded from the treatments. Therefore, further study of these types of fibroids and the factors affecting their treatability should be considered in the future.

\citet{mindjuk2015mri} conducted statistical univariate and multivariate analyses on multiple clinical parameters affecting the treatment outcome in uterine fibroids. In the univariate analysis it was found that subserosal fibroid location, fibroid septation, fibroid volume, T2-weighted (i.e., equal to Funaki type) and CE-T1-weighted fat saturated signal intensities, and fibroid distance from spine and skin were the most significant parameters correlating with the immediate post-treatment NPV ratio. In the multivariate analysis the fibroid volume and T2w signal intensity were not found to be significant factors. It was also shown that the NPV ratio is the only statistically significant variable for predicting the risk of retreatment, and thus, it can serve as an indicator of clinical success. In this study, subserosal fibroid location placed only 32th in the aggregate rankings, but fibroid volume, fibroid distance from the skin surface, and Funaki types I and III were all in top ten.

Keserci and Duc~\citep{keserci2018magnetic} studied the role of MRI parameters in predicting the treatment outcome of HIFU therapy in uterine fibroids with NPV ratio of at least 90\%. It was shown with a multivariate analysis that the thickness of the subcutaneous fat layer, T2-weighted signal intensity ratio of fibroid to myometrium (i.e., Funaki type I/II or III) and MRI perfusion parameters had significant correlation with the immediate post-treatment NPV ratio of at least 90\%. MRI perfusion parameters were not included in this study, but subcutaneous fat layer thickness, and again, Funaki types I and III were placed in top ten in the aggregate rankings.

The novelty of this study was to employ a wide selection of different filter-based feature selection methods to evaluate the information content of multiple clinical parameters in classifying the treatment outcome. In addition, a machine learning model (SVC) was used for the first time to take the advantage of these features in predicting the treatment outcome in three separate classes defined by the immediate post-treatment NPV ratio. The results showed that it is important to select the optimal feature set in order to achieve the best prediction performance with the chosen classification model.

Similar framework was utilised in a study by \citet{parmar2015machine} who used 14 different filter-based feature selection methods to choose radiomic features for predicting the two-year survival rate of lung cancer patients. In addition, they evaluated the performance of 12 classification methods in predicting the survival rate. It was found out that for the given problem Wilcoxon feature selection method had the highest performance whereas CHI2 and CIFE performed the worst. For classification methods, random forest showed the highest predictive performance whereas decision tree had the lowest. It was concluded that the choice of classification model is the most dominant source of performance variation. In this study the best performing feature selection methods were JMI and MIM while FISH and FSCR performed the worst with ten highest-ranking features.

The algorithm pipeline used in this study was a modification of the one used by \citet{deist2018machine} who compared different classifiers in predicting the outcome of (chemo)radiotherapy. However, they did not employ separate filter-based feature selection methods because all the classifiers had built-in dimensionality reduction (i.e., embedded or wrapper feature selection). Random forest and elastic net logistic regression models yielded the highest discriminative performance among the studied 12 datasets, but there was no single best classifier across all datasets. 

In this study only a single classification model (SVC) was used because of its omission of built-in dimensionality reduction and robust performance with small but clean datasets. The classification performance could potentially have been further improved by using balanced classes and/or different classifier. However, the aim of this study was not to achieve the best possible classification performance, but to find out the most important and informative features from the studied clinical parameters. Therefore, the results of this study will act as an important reference for future research using different machine learning models in predicting HIFU treatment outcome in uterine fibroids.

The limitations of the study include the omission quantitative perfusion values (MRI-T1~\citep{keserci2017role} or contrast-enhanced US~\citep{wang2016influence}) from the analysis, which would likely rank high in terms of their correlation with the treatment outcome. T1-perfusion images are not included in the standard set of images acquired in the patient screening scans, and therefore, these values were not analysed due to the scarcity of available data. Therefore, perfusion values should be considered in potential future studies.

Ultimately the aim is to create a machine learning model which could accurately predict the HIFU therapy outcome in uterine fibroids with as few clinical parameters as possible without compromising the performance of the model, and therefore, the inclusion of all the possible features in the final model is neither desired nor purposeful. Furthermore, these parameters should be easily obtainable in order to streamline the patient screening workflow. Lastly, the final model should be fast and simple enough to use in order to be adopted in clinical practice. The prediction model would not replace the radiologist or gynaecologist in the screening process, but rather would assist the patient selection by giving a quantitative and objective estimate of the treatment outcome based on the pre-treatment parameters.

\pagebreak

\bibliographystyle{UMB-elsarticle-harvbib}
\bibliography{arXiv_scientific_reports}


\end{document}